\documentclass[trackchanges]{aastex701}
\usepackage{CJK}
\usepackage{xcolor}

\newcommand\latex{La\TeX}

\begin{document}

\title{JWST reveals how black holes are fed: kiloparsec-scale multiphase filaments feed sub-kiloparsec circumnuclear disks}

\shorttitle{JWST Observations of NGC~4696}
\shortauthors{Hlavacek-Larrondo et al.}

\begin{CJK}{UTF8}{}
\CJKfamily{mj}

\correspondingauthor{Julie Hlavacek-Larrondo}
\email{j.larrondo@umontreal.ca}

\author[0000-0001-7271-7340]{Julie Hlavacek-Larrondo}
\affiliation{D\'{e}partement de Physique, Universit\'{e} de Montr\'{e}al, Succ. Centre-Ville, Montr\'{e}al, Qu\'{e}bec, H3C 3J7, Canada}
\email{...}  

\author[0000-0002-3173-1098]{Hyunseop Choi (최현섭)}
\affiliation{Department of Astronomy, University of Michigan, 1085 S. University, Ann Arbor, MI 48109, USA}
\email{...}  

\author[0000-0002-3680-5420]{Minghao Guo (郭明浩)}
\affiliation{Department of Astrophysical Sciences, Princeton University, Princeton, NJ 08540, USA}
\email{...}  

\author[0009-0008-7156-4678]{Mathieu Marquis}
\affiliation{D\'{e}partement de Physique, Universit\'{e} de Montr\'{e}al, Succ. Centre-Ville,
Montr\'{e}al, Qu\'{e}bec, H3C 3J7, Canada}
\email{...}  

\author[]{Olivia Pereira}
\affiliation{D\'{e}partement de Physique, Universit\'{e} de Montr\'{e}al, Succ. Centre-Ville,
Montr\'{e}al, Qu\'{e}bec, H3C 3J7, Canada}
\email{...}  

\author[0000-0002-3514-0383]{G. Mark Voit}
\affiliation{Department of Physics and Astronomy, Michigan State University,
East Lansing, MI 48824, USA}
\email{}

\author[0000-0003-0475-9375]{Lo\"ic Albert}
\affiliation{Institut Trottier de recherche sur les exoplan\`etes and D\'epartement de Physique, Universit\'e de Montr\'eal, 1375 Avenue Th\'er\`ese-Lavoie-Roux, Montr\'eal, QC H2V 0B3, Canada}
\email{placeholder}

\author{Jorge Barrera-Ballesteros}
\affiliation{Instituto de Astronom\'{i}a, Universidad Nacional Aut\'{o}noma de M\'{e}xico,
A.P. 70-264, 04510 CDMX, M\'{e}xico}
\email{}

\author{Rebecca E. A. Canning}
\affiliation{Institute of Cosmology and Gravitation, University of Portsmouth,
Dennis Sciama Building, Portsmouth, PO1 3FX, UK}
\email{}

\author{Francesco D'Eugenio}
\affiliation{Kavli Institute for Cosmology, University of Cambridge,
Madingley Road, Cambridge, CB3 0HA, UK}
\affiliation{Cavendish Laboratory, University of Cambridge,
19 JJ Thomson Avenue, Cambridge, CB3 0HE, UK}
\email{}

\author{Megan Donahue}
\affiliation{Department of Physics and Astronomy, Michigan State University,
East Lansing, MI 48824, USA}
\email{}

\author{Andrew C. Fabian}
\affiliation{Institute of Astronomy, University of Cambridge,
Madingley Road, Cambridge, CB3 0HA, UK}
\email{}

\author{Gary J. Ferland}
\affiliation{Department of Physics and Astronomy, University of Kentucky,
505 Rose Street, Lexington, KY 40506, USA}
\email{}

\author[0000-0001-8608-0408]{John S. Gallagher}
\affiliation{Department of Physics and Astronomy, Macalester College,
1600 Grand Avenue, Saint Paul, MN 55105, USA}
\email{}

\author[0000-0002-7326-5793]{Marie-Lou Gendron-Marsolais}
\affiliation{D\'{e}partement de Physique, de G\'{e}nie Physique et d'Optique,
Universit\'{e} Laval, Qu\'{e}bec, QC G1V 0A6, Canada}
\email{}

\author{Pierre Guillard}
\affiliation{Sorbonne Universit\'{e}, CNRS, UMR 7095, Institut d'Astrophysique de Paris,
98bis bd Arago, 75014 Paris, France}
\email{}

\author{Nina Hatch}
\affiliation{School of Physics and Astronomy, University of Nottingham,
University Park, Nottingham, NG7 2RD, UK}
\email{}


\author{Ralf Kotulla}
\affiliation{Department of Astronomy, University of Wisconsin--Madison,
475 N. Charter Street, Madison, WI 53706, USA}
\email{}

\author[0000-0001-5262-6150]{Yuan Li}
\affiliation{Department of Astronomy, University of Massachusetts,
Amherst, MA 01003, USA}
\email{}

\author{Roberto Maiolino}
\affiliation{Kavli Institute for Cosmology, University of Cambridge,
Madingley Road, Cambridge, CB3 0HA, UK}
\affiliation{Cavendish Laboratory, University of Cambridge,
19 JJ Thomson Avenue, Cambridge, CB3 0HE, UK}
\affiliation{Department of Physics and Astronomy, University College London,
Gower Street, London WC1E 6BT, UK}
\email{}

\author{Allison Man}
\affiliation{Department of Physics \& Astronomy, University of British Columbia,
6224 Agricultural Road, Vancouver, BC V6T 1Z1, Canada}
\email{}

\author{Michael A. McDonald}
\affiliation{Kavli Institute for Astrophysics and Space Research,
Massachusetts Institute of Technology,
77 Massachusetts Avenue, Cambridge, MA 02139, USA}
\email{}

\author{B. R. McNamara}
\affiliation{Department of Physics and Astronomy, University of Waterloo,
200 University Avenue West, Waterloo, ON N2L 3G1, Canada}
\email{}

\author{Valeria Olivares}
\affiliation{Departamento de F\'{i}sica, Universidad de Santiago de Chile,
Av. Victor Jara 3659, Santiago 9170124, Chile}
\affiliation{Center for Interdisciplinary Research in Astrophysics and Space Exploration (CIRAS),
Universidad de Santiago de Chile, Santiago 9170124, Chile}
\email{}

\author[0009-0003-0932-2487]{Marine Prunier}
\affiliation{D\'{e}partement de Physique, Universit\'{e} de Montr\'{e}al,
Succ. Centre-Ville, Montr\'{e}al, Qu\'{e}bec, H3C 3J7, Canada}
\affiliation{Max-Planck-Institut f{\"u}r Astronomie, K{\"o}nigstuhl 17, D-69117 Heidelberg, Germany}
\email{}

\author{Michael Reefe}
\affiliation{Kavli Institute for Astrophysics and Space Research,
Massachusetts Institute of Technology,
77 Massachusetts Avenue, Cambridge, MA 02139, USA}
\email{}

\author[0000-0002-1510-4860]{Christopher S. Reynolds}
\affiliation{Department of Astronomy, University of Maryland,
College Park, MD 20742-2421, USA}
\affiliation{Joint Space Science Institute (JSI), University of Maryland, College Park, MD 20742-2421, USA}
\email{}

\author{Carter Rhea}
\affiliation{Dragonfly Focused Research Organization, 150 Washington Avenue, Santa Fe, 87501, NM, USA}
\email{}

\author{Annabelle Richard-Laferri\`{e}re}
\affiliation{Institute of Astronomy, University of Cambridge,
Madingley Road, Cambridge, CB3 0HA, UK}
\email{}

\author{Helen R. Russell}
\affiliation{School of Physics \& Astronomy, University of Nottingham,
University Park, Nottingham, NG7 2RD, UK}
\email{}

\author{Philippe Salom\'{e}}
\affiliation{LUX, Observatoire de Paris, Université PSL, Sorbonne Université, CNRS, 75014 Paris, France}
\email{}

\author[0000-0001-5880-0703]{Ming Sun}
\affiliation{Department of Physics and Astronomy, University of Alabama in Huntsville,
301 Sparkman Drive, Huntsville, AL 35899, USA}
\email{}

\author[0000-0001-8176-7665]{Prathamesh Tamhane}
\affiliation{Department of Physics and Astronomy, University of Alabama in Huntsville,
301 Sparkman Drive, Huntsville, AL 35899, USA}
\email{}

\author{Gregory Taylor}
\affiliation{Department of Physics and Astronomy, University of New Mexico,
Albuquerque, NM 87131, USA}
\email{}

\author{Auriane Thilloy}
\affiliation{D\'{e}partement de Physique, Universit\'{e} de Montr\'{e}al,
Succ. Centre-Ville, Montr\'{e}al, Qu\'{e}bec, H3C 3J7, Canada}
\email{}

\author[0000-0002-5445-5401]{Grant R.~Tremblay}
\affiliation{Harvard-Smithsonian Center for Astrophysics,
60 Garden Street, Cambridge, MA, USA}
\email{}

\author{Benjamin Vigneron}
\affiliation{D\'{e}partement de Physique, Universit\'{e} de Montr\'{e}al,
Succ. Centre-Ville, Montr\'{e}al, Qu\'{e}bec, H3C 3J7, Canada}
\email{}

\author[0000-0002-6413-4142]{Stephen A. Walker}
\affiliation{Department of Physics and Astronomy, University of Alabama in Huntsville,
301 Sparkman Drive, Huntsville, AL 35899, USA}
\email{}


\begin{abstract}
The Centaurus cluster is one of the most important archetypes of radio-mode AGN feedback, with its central galaxy, NGC~4696, launching powerful jets that inflate X-ray cavities and regulate cooling and star formation. NGC~4696 lies within a spectacular multiphase nebula of filaments extending over tens of kiloparsecs and spanning six decades in temperature, from hot ($10^8$~K) X-ray-emitting plasma to cold molecular gas. Owing to its proximity, \textit{Hubble Space Telescope} H$\alpha$ imaging reveals a striking S-shaped ionized-gas swirl within the black hole’s sphere of influence---the first such structure identified in a cluster core \citep{Fabian2016}. Here we present the first JWST observations of NGC~4696 with NIRSpec, probing the inner 618~pc~$\times$~618~pc at 10~pc resolution. These data reveal that the ionized swirl is a rotating, multiphase circumnuclear disk (CND) physically and kinematically connected to the larger-scale filamentary network. This provides the long-sought missing link between kiloparsec-scale cooling flows and black hole accretion on $<100$ pc scales. Strikingly, the observed morphology and kinematics are reproduced by tailored magnetohydrodynamic simulations, in which filamentary gas condenses from the hot atmosphere, loses angular momentum, and feeds a rotating CND that mediates accretion onto the black hole. A similar structure in NGC~1275, the Perseus cluster’s central galaxy \citep{oosterloo2023}, together with our results on NGC~4696---two prototypical radio-mode AGN feedback systems---points to a common mechanism: multiphase filaments transport gas from cluster scales down to the vicinity of the black hole via a CND, closing the AGN feedback loop and establishing a physically grounded framework for self-regulated galaxy evolution.
\end{abstract}


\section{Introduction} \label{sec:intro}

The best place to study supermassive black hole (SMBH) feedback is in the hot atmospheres of galaxy clusters, which host the most massive black holes and where the impact of these black holes can be directly observed. The X-ray-bright intracluster medium (ICM) allows us to trace how energy released by the central SMBH, through powerful relativistic jets, injects mechanical energy that offsets radiative cooling and suppresses star formation \citep{McNamara2012,Fabian2012,HL2022}. Remarkably, this feedback appears to be self-regulated despite the vast scale difference between the black hole and its host galaxy. 

Several closely-related theoretical frameworks---including the precipitation model \citep[e.g.][]{Voit2015-N}, cold feedback \citep{Pizzolato2005}, chaotic cold accretion \citep{Gaspari2013}, and stimulated feedback \citep[e.g.,][]{McNamara2016,Li2014,Revaz2008}---attribute this regulation to multiphase gas cooling and inflow, emphasizing the need to trace gas flows continuously from kiloparsec scales down to the vicinity of the black hole \citep{Hopkins2010,Hopkins2011,Li2014,Guo2023,Guo2024}. A key prediction of these frameworks is that some cooling clouds must have sufficiently low angular momentum to reach the SMBH \citep{Pounds2018}, implying intrinsically disordered inflow near the nucleus \citep[e.g.,][]{Gaspari2013,Gaspari2018}. Observational support has largely come from molecular gas seen in absorption against compact radio cores \citep{David2014,Tremblay2016,Rose2019a,Rose2019b}, probing kinematics consistent with inflow within the inner few hundred parsecs \citep[e.g.][]{Rose2019b,Nagai2019}, but only along narrow sight-lines, providing radial velocities rather than a full spatial view of the flow.
 
Here, we focus on the Centaurus cluster of galaxies located at $z=0.01003$ \citep[luminosity distance 43.3 Mpc; $206$ pc arcsec$^{-1}$][]{XRISMNGC4696}, one of the few systems where gas flows can be spatially resolved down to scales comparable to the black hole’s sphere of influence, enabling direct tests of accretion models while providing a uniquely detailed view of the central gas compared to more distant clusters such as Perseus \citep[e.g.][and references therein]{Minkowski1957,Conselice2001} or Abell 2597 \citep[e.g.][]{Tremblay2016}.

The central galaxy, NGC~4696, is embedded in a giant multiphase nebula of filaments (see Fig. \ref{fig:hst_jwst_paa})---typical of cluster cores spanning $\sim6$ decades in temperature, from X-ray gas at $10^7$--$10^8$ K to $\sim30$ K molecular gas \citep[e.g.][]{Crawford2005,Olivares2019,Hamer2019}. These narrow ($\sim70$ pc) filaments, extending over several kiloparsecs, likely form through a combination of in situ condensation (precipitation) and uplifted low-entropy gas becoming thermally unstable (stimulated feedback). The ionization of the gas is not dominated by star formation, but instead reflects a close coupling to the ICM and active galactic nucleus (AGN) feedback processes, with contributions from atomic, ionized, and molecular phases (e.g., H$\alpha$, low-ionization lines, [\ion{C}{2}], H$_2$, CO). Proposed heating mechanisms include magnetic field reconnection \citep[e.g.][]{Churazov2013}, excitation of turbulent mixing layers \citep[e.g.][]{Crawford1992}, as well as heating due to collisions from ICM energetic particles \citep[e.g.][]{Ferland2009}.

Deep HST H$\alpha$ images (Fig. \ref{fig:hst_jwst_paa}) show that the filaments in NGC~4696 may be magnetically supported, similar to those in Perseus \citep{Fabian2008, Fabian2016,Tamhane2026}, and reveal a nuclear S-shaped swirl ($\sim1''$, 206 pc) whose emission peaks near the center and lies largely within or just outside the black hole sphere of influence \citep[see][]{Fabian2016}. However, HST imaging provides only morphology, not kinematics, which are essential to trace gas flows and test accretion models.

We therefore present the first James Webb Space Telescope (JWST) observations of NGC~4696 obtained with the Near-Infrared Spectrograph (NIRSpec) instrument \citep{boker2022}, aimed at resolving the kinematics of the nuclear swirl within the black hole’s sphere of influence. We show that the ionized swirl is a circumnuclear rotating disk (CND) directly connected to the outer filaments, providing the missing link between kiloparsec-scale cooling and sub-kpc accretion. Section \ref{obs} describes the observations, and Section \ref{sims} presents tailored three-dimensional magnetohydrodynamic (3DMHD) simulations of SMBH fueling. We present and discuss the results in Section \ref{res} and consider their implications in Section \ref{conc}. We adopt $H_0=70$ km s$^{-1}$ Mpc$^{-1}$, $\Omega_{\rm m}=0.3$, and $\Omega_{\rm \Lambda}=0.7$, with 206 pc per~\arcsec\ at $z=0.01003$. Errors are 1$\sigma$ unless stated otherwise.

\section{Observations and Analysis} \label{obs}

NGC~4696 was observed on 4 June 2025 with the JWST/NIRSpec Integral Field Unit (IFU; program 5354, PI: Hlavacek-Larrondo) using the F170LP/G235H configuration centered on the core, covering a $3\farcs0\times3\farcs0$ Field Of View (FOV; 618 pc $\times$ 618 pc). The setup provides near-continuous coverage from $\sim$1.66-3.17 $\mathrm{\mu m}$, with a $\sim$2.40-2.56 $\mathrm{\mu m}$ gap between detectors. The G235H grating has $R\sim2700$ (velocity resolution $\sim110\ \mathrm{km\ s^{-1}}$), with a total on-source exposure time of 7.7 hours.

The data were reduced with JWST Build 12.2 using CRDS context 1464 pmap and standard STScI pipelines \citep{Bushouse2019}, with \texttt{snowblind}\footnote{\href{https://github.com/mpi-astronomy/snowblind}{https://github.com/mpi-astronomy/snowblind}} replacing the default algorithm to mitigate snowball/shower artifacts and \texttt{nsclean} applied for $1/f$ noise subtraction \citep{Rauscher2024}. Exposures were drizzle-combined into a final cube sampled at 0.05'' per spaxel ($\sim10$ pc), followed by additional outlier rejection to remove any remaining bad pixels. To correct PSF undersampling-induced spectral undulations near the nucleus, we applied mild smoothing with a 2.5 spaxel circular aperture ($\sim0.12$'', comparable to the $\sigma$ of the PSF at the Pa$\alpha$ wavelength; e.g., \citealt{Bentz2025,Liu2024}), effectively removing oscillation artifacts with minimal impact on emission lines.

For spectral analysis, we used a modified \texttt{q3dfit} \citep{Rupke2023} to model the continuum and line emissions, adopting a single empirical stellar-continuum template from a line-free region of the NIRSpec FOV that was scaled during fitting rather than modeled per spaxel. Here, we only focus on the Pa$\alpha$ 1.87~$\mathrm{\mu m}$ line as the primary tracer of warm ($10^4$ K) ionized gas distribution and kinematics, given its strength and diagnostic power within our bandpass. A full spectral analysis, including stellar continuum modeling and emission-line fitting across the entire observed bandpass (e.g., cooler molecular $\rm H_2$ 1-0 ro-vibrational lines), will be presented in future work (Marquis et al. submitted; Pereira et al. in prep).

\begin{figure*}
\centering
\includegraphics[width = 0.6\textwidth]{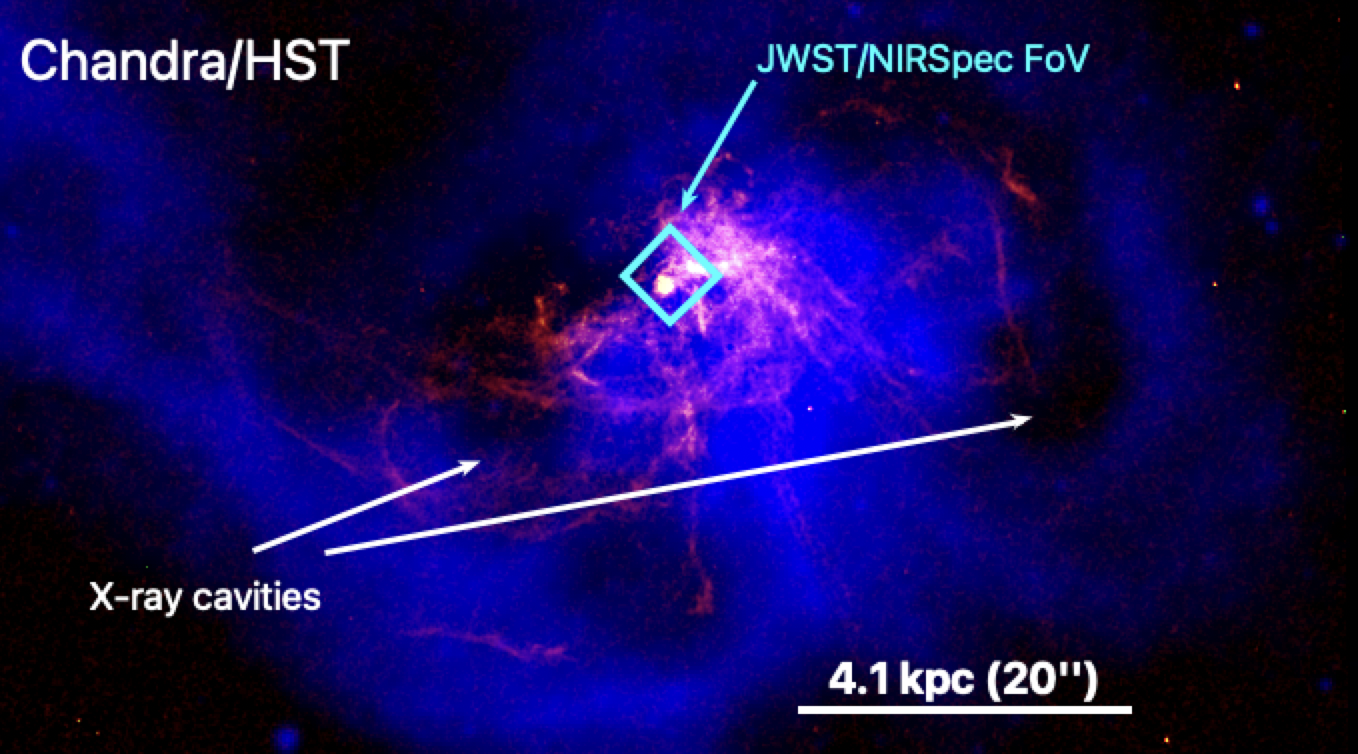}\vspace{0.2cm}

\includegraphics[width = 1\textwidth]{figs/paa_jwst_fig1_9_2d.png}
\caption{Top panel: Composite large-scale image of the Centaurus cluster. Chandra 0.5 to 7 keV image in blue. HST/WFC3 F665N (containing redshifted H$\alpha$), with the continuum estimated from F814W and subtracted, shown in red. The field of view (FoV) of JWST/NIRspec is shown with the cyan square. Middle and bottom panels: The central $760$ pc $\times~550$ pc of NGC~4696. Top-left: HST/WFC3 F665N (containing redshifted H$\alpha$), with the continuum estimated from F814W and subtracted, revealing the S-shaped ionized swirl.
The approximate edge of the assumed CND is shown in a red circle.
Top-right: same image in greyscale but with Pa$\alpha$ flux map (NIRSpec) overlaid.
Bottom-left/right: same with $v_{\rm 50}$ (centroid velocity) and $w_{\rm 80}$ (80\% line width) maps, respectively, derived from the NIRSpec cube (Section \ref{obs}).
We see a CND connected to a filamentary structure to the west. The AGN position \citep{Fabian2016} is marked (green circle), with the arrow in $v_{\rm 50}$ indicating the VLBA jet direction \citep{Taylor2006}; velocities are relative to the systemic of NGC~4696.}
\label{fig:hst_jwst_paa}
\end{figure*}

The line was fit with a single Gaussian profile.
To avoid fitting noise spikes or very broad non-astrophysical features,
we constrained the line width to $\sigma \lesssim 1000\ \mathrm{km\ s^{-1}}$ \citep[e.g.,][]{Vayner2023}.
We adopted a flux threshold of $1\times10^{-19}\ \mathrm{erg\ s^{-1}\ cm^{-2}}$ to identify robust emission-line detections ($\gg1\sigma$).
To further filter out any remaining noise spikes or spurious fits from the final results, we selected only spaxels with
velocity centroid ($v_{50}$) to the range $-300$ to $+400\ \mathrm{km\ s^{-1}}$ and
line width to $w_{80}\lesssim800\ \mathrm{km\ s^{-1}}$ (corresponding to the overall width containing 80\% of line flux; $\sigma \lesssim 300\ \mathrm{km\ s^{-1}}$),
except in the innermost nuclear region (CND; see the middle panel in Figure~\ref{fig:hst_jwst_paa}) where broader kinematic components may be physically expected.
These selection thresholds were carefully chosen to not erroneously remove any genuine detections; we verified that relaxing these constraints has no impact on our results.

The IFU astrometry can be uncertain; we therefore assigned a new WCS to the cube by matching the swirl observed in $HST$/WFC3 $F665N$ \citep{Fabian2016} to same swirl seen in the JWST Pa$\alpha$ line map, with a full multiwavelength astrometric comparison incorporating Pa$\alpha$ kinematics presented in Marquis et al. (submitted).


\section{Numerical Simulations} \label{sims}

\begin{figure*}
\centering
\includegraphics[width = 0.99\textwidth]{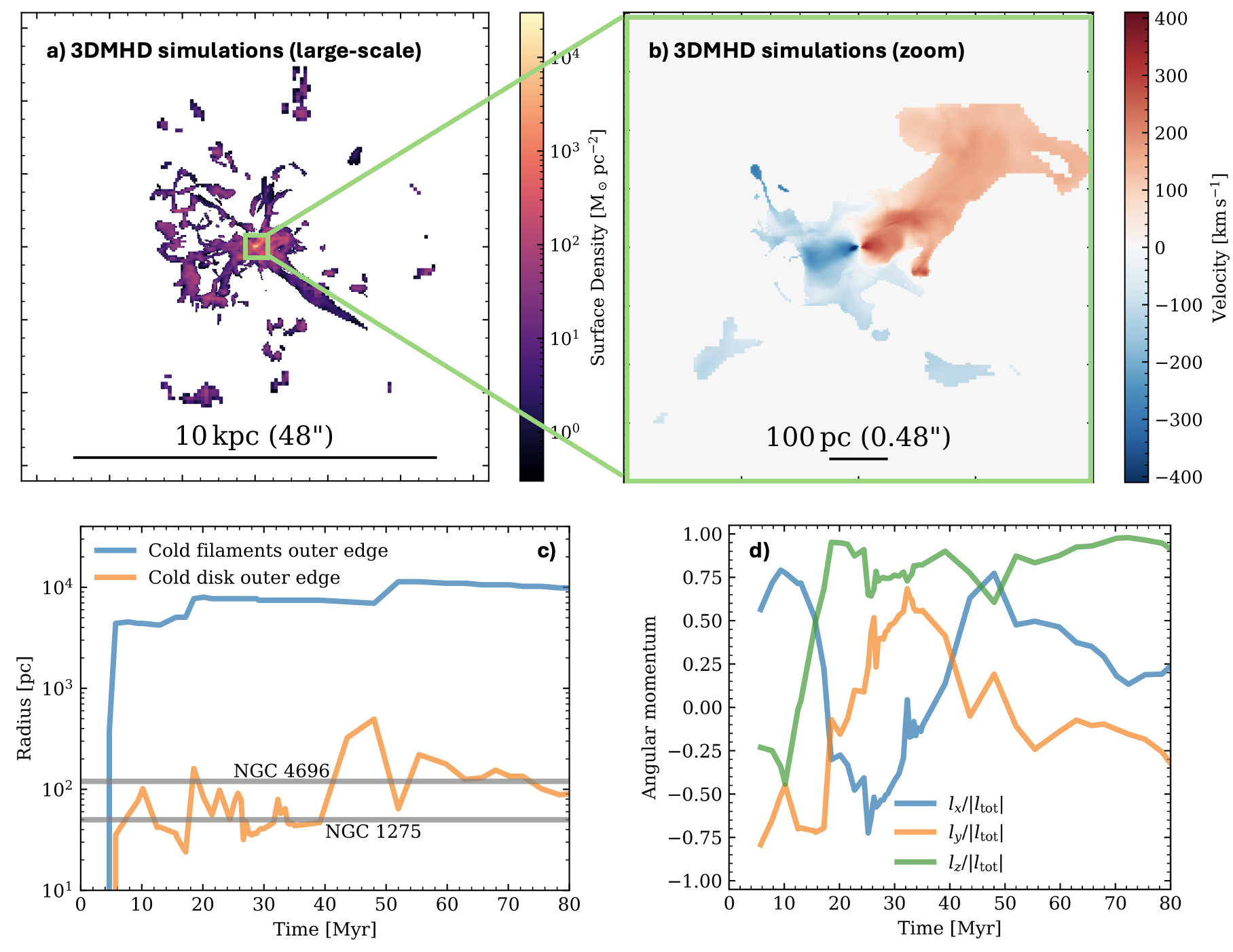}
\caption{Top panels: a) Snapshot of projected large-scale 3DMHD simulated filament network. b) Snapshot of 3DMHD simulated velocity field within a NIRSpec/IFU field of view, showing only cool gas with a surface density greater than $1\%$ of the maximum. Both snapshots shown in panels a) and b) corresponds to $t = 40$ Myr of the 3DMHD simulations. Bottom panels: c) Evolution of the outer radius of the kpc-scale filaments (blue) and the cool ($\sim10^{4-5}$ K) CND (orange) as a function of time. We define the CND size as the radius at which the spherically averaged angular momentum of the cool gas drops below 95\% of the Keplerian value. The filaments rapidly expand to kpc scales and remain relatively stable, while the CND exhibits variability, with episodic growth and contraction on $\sim$10-100 pc scales. The disc size of the CND in NGC 4696 (this work) and NGC 1275 \citep{oosterloo2023} are also shown. d) Same as Panel c), but tracing the evolution of the orientation of the CND as traced by its angular momentum direction along three different projections.} . 
\label{fig:simsobs}
\end{figure*}

To interpret the physical origin of the observed filament-CND connection and place it within a self-consistent framework for SMBH fueling, we compare our results to tailored 3DMHD simulations that capture the multiphase condensation and inflow of gas from kiloparsec to sub-parsec scales. More specifically, we perform 3DMHD simulations of a turbulent, cooling medium in an elliptical galaxy, following a setup similar to that of~\citet{Guo2024}, using the code AthenaK~\citet{Stone2026ApJS..283...27S}. To approximate the properties of NGC~4696, we adopt a spherically symmetric gravitational potential with a black hole mass of $M_\mathrm{BH}=10^9\,M_\odot$ \citep[see][for black hole mass estimates in NGC 4696]{Russell2013,Mezcua2018}, plus stellar and dark matter components both described by NFW profiles, with characteristic masses $3\times10^{10}\,M_\odot$ and $10^{12}\,M_\odot$, and scale radii of $1\,\mathrm{kpc}$ and $30\,\mathrm{kpc}$ respectively. The initial conditions are constructed by solving for hydrostatic equilibrium with a cored entropy profile, $K = K_0(1 + r/r_0)$, where $r_0 = 1\,\mathrm{kpc}$ and $K_0 = 2.3\,\mathrm{keV\,cm^2}$ with the density $n_0 = 0.2\,\mathrm{cm^{-3}}$ at $r_0$. Initial turbulence with velocity dispersion of $\sim 30-50\,\mathrm{km\,s^{-1}}$ and an entangled magnetic field with plasma-$\beta\simeq 100$ are imposed. The simulations adopt a cubic domain spanning $[-12,12]^3\,\mathrm{kpc}^3$ with a root grid of $128^3$ cells and 8 levels of mesh refinement. At the inner boundary, we apply a sink with radius $r_\mathrm{sink} = 6\,\mathrm{pc}$. We adopt optically thin bremsstrahlung and line cooling for solar metallicity. Details of the numerical methods are provided in \citet{Guo2023} and \citet{Guo2024}. The simulation is evolved for 80 Myr. We present one of the snapshots of these simulations in Fig. \ref{fig:simsobs}.

\section{Results and discussion}\label{res}

The central $\sim1''$ region of NGC~4696 hosts a striking S-shaped ionized structure near the black hole’s sphere of influence (Fig. \ref{fig:hst_jwst_paa}), first revealed by \emph{HST} \citep{Fabian2016}. Our new \emph{JWST}/NIRSpec observations resolve its kinematics for the first time, showing that the swirl is a CND physically connected to a multiphase filament extending west. This provides a direct link between kpc-scale filaments and gas dynamics within $\lesssim100$ pc. We discuss the properties of the CND, its connection to filamentary inflow, comparisons with 3DMHD simulations, and implications for AGN fueling and feedback.

\subsection{Circumnuclear rotating disk}\label{CDN}

The black hole’s sphere of influence can be estimated as $r_{\rm inf}=GM_{\rm BH}/\sigma^2$. Using $K$-band scaling relations between black hole mass and host galaxy properties, \citet{Russell2013} inferred $M_{\rm BH}\sim10^9~M_\odot$ for NGC 4696 (consistent with \citealt{Mezcua2018}), implying $r_{\rm inf}\sim70$ pc ($\sim0.3''$). We stress that this is a very rough estimate and that detailed dynamical modelling over the CND would be required to constrain more reliably the sphere of influence of the black hole. If correct, then the NIRSpec/IFU data resolve gas dynamics across this region, sampling the central $\sim1''$ swirl at $\sim0.1''$ resolution and enabling mapping across and beyond the sphere of influence.

In the middle-left panel of Fig. \ref{fig:hst_jwst_paa}, we show the S-shaped swirl seen in the HST H$\alpha$ image. The middle-right panel presents the JWST Pa$\alpha$ flux map, where this structure is also clearly detected. The Pa$\alpha$ kinematics (bottom panels) further reveal that this swirl is in fact a compact CND with a radius of $\sim0.6''$ (or $\sim$120 pc) centered around the AGN, with velocities ranging from $\sim-200$ to $+400\ \mathrm{km\ s^{-1}}$. The velocity dispersion ($w_{80}$) rises sharply near the AGN and the jet hotspot, reaching values up to $\sim1700\ \mathrm{km\ s^{-1}}$, consistent with gas moving in the deep potential of the central SMBH. Connected to the disk is an ionized filament in Pa$\alpha$, with a width of $\sim105$ pc and a length of $\sim350$ pc (although the length most likely extends beyond the NIRSpec/IFU field of view). Near the interface between this ionized filament and the edge of the CND, $w_{80}$ reaches $\sim300\ \mathrm{km\ s^{-1}}$. Moving along the filament from east to west, $w_{80}$ gradually declines to $\sim130\ \mathrm{km\ s^{-1}}$. Over the same region, the velocity decreases smoothly from $\sim100\ \mathrm{km\ s^{-1}}$ near the CND edge toward the systemic velocity. This suggests that the gas is more turbulent where the filament joins the CND, potentially reflecting ongoing accretion onto the (CND; see Section \ref{feedbackloop}).

The CND we observe is detected across multiple emission lines with the NIRSpec data---not only Pa$\alpha$ (tracing $\sim10^4$ K gas), but also several H$_2$ transitions (notably $\rm H_2$ 1-0 S(1) and $\rm H_2$ 1-0 S(3)) that trace colder gas (few $10^2$--$10^3$ K). The CND is therefore multiphase. Here, we focus only on the Pa$\alpha$ line and discuss the properties of the other lines in subsequent papers (Marquis et al. submitted; Pereira et al. in prep). Indeed, including the H$_2$ transitions would require detailed modelling of the stellar continuum, which is beyond the scope of this paper.

XRISM X-ray spectroscopy of the Centaurus cluster \citep{XRISMNGC4696} shows that the hot intracluster medium flows along the line of sight with velocities of $\sim130$-$310$ km s$^{-1}$ within the inner 30 kpc, consistent with gas sloshing around the cluster center. The velocity dispersion remains low ($<120$ km s$^{-1}$) even within the inner 10 kpc, indicating a limited impact of the AGN on the surrounding gas motions. Overall, the hot gas structure appears dominated by sloshing \citep{Sanders2016,XRISMNGC4696}. The Centaurus cluster consists of two main subclusters, Cen 30 and Cen 45 \citep{XRISMNGC4696}, with NGC 4696 being the central galaxy of Cen 30. \citet{XRISMNGC4696} suggest that the observed sloshing is associated with oscillatory motions of NGC 4696 within the cluster potential, likely triggered by the interaction between Cen 30 and Cen 45. Such motions can induce relative gas flows between the galaxy and the surrounding intracluster medium, altering ram-pressure confinement and influencing gas transport and AGN feeding. Thus, sloshing motions may strongly impact both the central galaxy and its AGN.

On kpc scales, the radio jets carve out large X-ray cavities as seen with the $Chandra$ X-ray Observatory, extend to a radius of $\sim5$ kpc, and are aligned roughly east-west \citep[see Fig. \ref{fig:hst_jwst_paa}, as well as][]{Sanders2016}. On sub-kpc scales, however, \citet{Taylor2006} showed that the radio jet appears to align roughly north-south (see middle and bottom panels of Fig. \ref{fig:hst_jwst_paa}). The radio jet therefore appears to change substantially in direction from kpc to sub-kpc scales, perhaps due to the strong sloshing occurring in the cluster. What is particularly interesting is that the sub-kpc scale radio jet appears roughly perpendicular to the CND, with the CND aligned predominantly in the east-west direction and the radio jet north-south. 

This is not the first time a sub-kpc CND has been discovered in a central cluster galaxy. A similar structure was reported in NGC 1275, the central galaxy of the Perseus cluster. It was first identified by \citet{Wilman2005} using the imaging spectrograph UIST (UKIRT Imaging Spectrometer) on UKIRT, and subsequently detected with the Near-Infrared Integral Field Spectrograph (NIFS) on Gemini North through the H$_2$ 1-0 S(1) and [\ion{Fe}{2}] 1.644 $\mu$m lines \citep{Scha2013}. The latter study noted that the gas kinematics in the outer parts of the CND appeared disturbed, similar to what we observe in NGC~4696, and interpreted this as possible evidence for gas accreting onto the disk.

The CND in NGC 1275 has a radius of approximately 50 pc and appears to be oriented roughly perpendicular to the radio jets \citep{Wilman2005,Scha2013,Nagai2019,oosterloo2023}. NGC~4696 shows a similar configuration: the CND has an east--west velocity gradient, with approaching (blueshifted) gas to one side and receding (redshifted) gas to the other, while the radio jets are oriented roughly north to south. However, the CND in NGC~4696 is approximately twice as large ($r\sim120$ pc) as that in NGC~1275. The velocity varies by roughly 600 km s$^{-1}$ across the CND in NGC~4696, comparable to what is observed in NGC 1275 \citep{oosterloo2023}. \citet{Nagai2019} imaged the CND in NGC 1275 at cold temperatures using CO(2-1), HCN(3-2), and HCO$^+$(3-2) with ALMA, finding a mass of $\sim10^8$ M$_\odot$ within the CND, though limited spatial resolution and artefacts restricted the analysis. 

In a broader context, \citet{Russell2019} analyzed ALMA observations of molecular gas in twelve central cluster galaxies, revealing massive cold gas reservoirs ($10^{9}$--$10^{11}$M$_{\odot}$), often in extended filamentary structures spanning tens of kiloparsecs and associated with AGN-inflated bubbles. These results support a scenario in which uplifted low-entropy X-ray gas becomes thermally unstable and cools into molecular filaments cycling through AGN-regulated feedback. \citet{Russell2019} also identified rare kpc-scale rotating disks (e.g., Phoenix, Abell 262, Hydra A), though limited spatial resolution prevented resolving sub-kpc CNDs and parsec-scale structures such as those seen in NGC~1275 and NGC~4696 \citep{Scha2013,Nagai2019,oosterloo2023}. Their prevalence and connection to kpc-scale disks therefore remain uncertain.

\subsection{Closing the feedback loop: filaments}\label{feedbackloop}

By reprocessing ALMA CO(2--1) data of NGC~1275, \citet{oosterloo2023} improved the resolution and mitigated imaging artefacts, allowing the molecular filaments to be decomposed into thinner, kinematically distinct structures and more clearly separated from the CND, although still with some overlap between filaments. They found that the filaments converge toward the CND, suggesting either inflow or outflow, and argued that the flow is predominantly inward. Their work highlights the importance of high-resolution observations for resolving the connection between kpc-scale filaments and the CND. We find evidence for a similar picture in NGC~4696. However, unlike NGC~1275, NGC~4696 does not exhibit multiple overlapping filaments within the central $618\,\mathrm{pc}\times618\,\mathrm{pc}$ region, making the connection between the filament and the CND much more clearly visible. We argue that the flow is also predominantly inward in NGC~4696, based on the same lines of evidence as \citet{oosterloo2023}.

First, \citet{oosterloo2023} showed that gas surrounding the CND in NGC 1275 is organized east-west, inconsistent with a jet-driven outflow given the north-south orientation of the radio jets. A similar configuration is observed in NGC~4696 (Fig. \ref{fig:hst_jwst_paa}): if the western filament were an outflow, it would be expected to align with the radio jet axis, essentially north to south as seen in VLBA observations on $\sim$30 pc scales and in 8.4 GHz VLA data on $\sim$200 pc scales \citep[see][]{Taylor2006}. Instead, it extends toward the west, indicating a different origin.

Second, the CND in NGC 1275 contains a fast-rotating component at its center, surrounded by a slightly larger disk-like structure with less regular kinematics, consistent with gas accreting onto the disk. We see a similar feature in NGC~4696, in which the velocity dispersion as traced with $w_{80}$ increases in the outskirts of our CND, notably to the west of the CND (see Fig. \ref{fig:hst_jwst_paa}). The Pa$\alpha$ velocity field shown in Fig. \ref{fig:hst_jwst_paa} also reveals positive (redshifted) velocity ionized gas on the east side of the CND, which may also indicate disordered infall and active accretion onto the CND.

\begin{figure*}
\centering
\includegraphics[width = 1\textwidth]{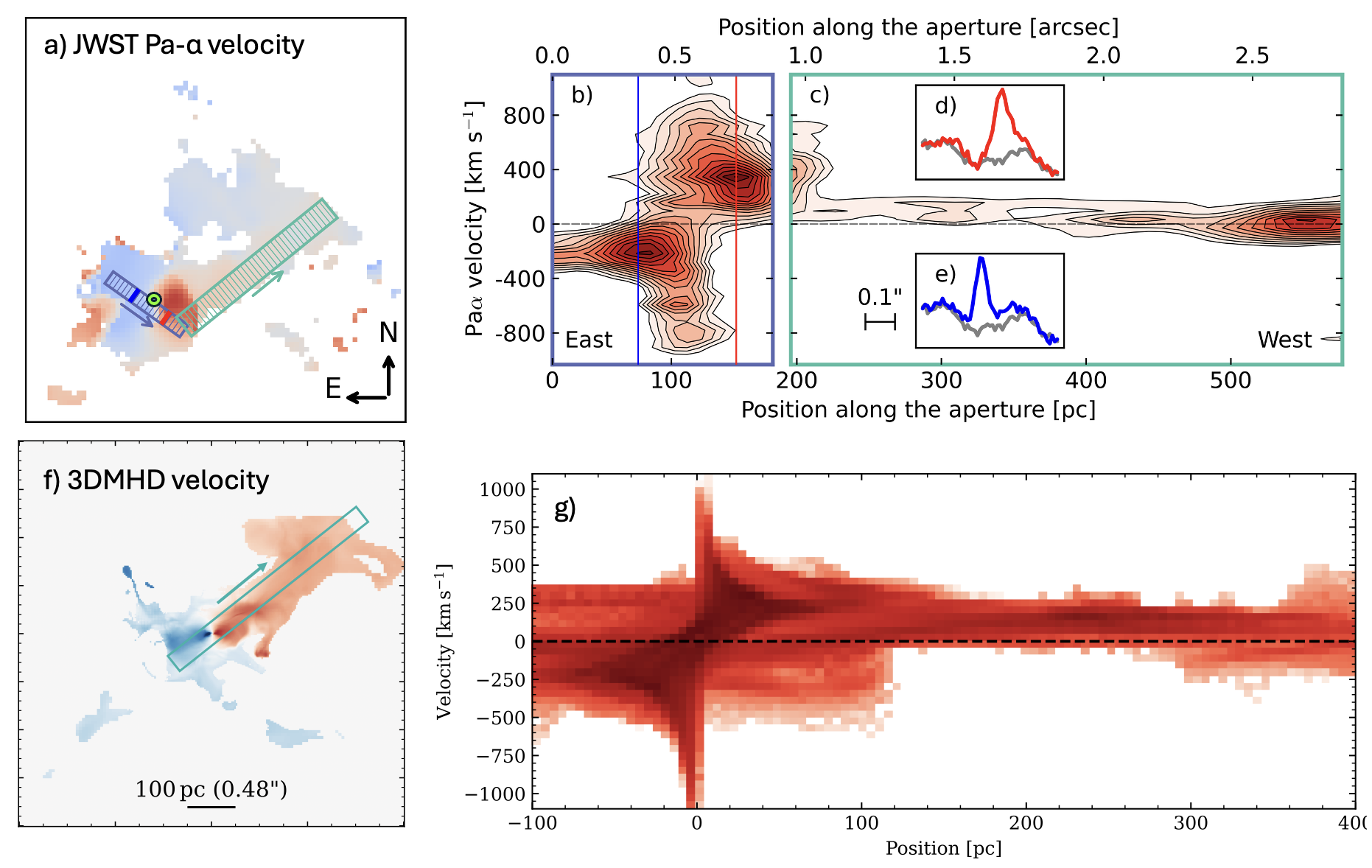}
\caption{Position-velocity (PV) diagrams. (a) $v_{\rm 50}$ map with slit positions and extraction directions \citep[green circle marks AGN position from][]{Fabian2016}. (b) PV diagram along the blue slit (CND), showing velocities rapidly declining to systemic on both sides. (c) PV diagram along the green slit, highlighting the filament’s kinematic connection to the CND. Velocities are relative to systemic (dashed horizontal line); contours start at $3\sigma$ and increase by $1\sigma$. The scale bar shows the NIRSpec/IFU spatial resolution. (d,e) Example Pa$\alpha$ spectra from the receding (red) and approaching (blue) sides, extracted from the regions and positions marked in (a,b). The continuum is shown with the grey curves. (f) Snapshot of 3DMHD simulated velocity field within a NIRSpec/IFU field of view, same as in Fig. \ref{fig:simsobs} with slit position shown with the green rectangle. (g) PV diagram along the green slit, highlighting the filament’s kinematic connection to the 3DMHD CND. }
\label{fig:pv}
\end{figure*}

Third, the position-velocity (PV) diagram of NGC 1275 shows that the outer CND is not only spatially connected to an eastern filament, but also kinematically linked, suggesting it feeds the disk and drives the more disturbed kinematics observed in its outer regions of the CND. We observe a similar behavior in the PV diagram of NGC~4696, shown in Fig.~\ref{fig:pv}. Here, PV diagrams were generated from the observations using the \texttt{pvextractor} Python package \citep{ginsburg_radio_2015}, which averages spectra along spatial bins and plots them as a function of position along an aperture. The plotted contour levels are determined relative to the noise ($\sigma_{\mathrm{rms}}$) sampled in a wavelength range close to the Pa$\alpha$ emission and averaged over all non-edge spaxels. The first aperture (blue) was chosen such that it passes through the kinematic center of the CND and along the maximum gradient in the rotation. The kinematic center and maximum gradient are detailed in Marquis et al. (submitted). The position of the second aperture (green) was selected to follow the filament and connect to the first slit, enabling us to probe the kinematic link between the two slits. We note that changing the orientation of the slit does not significantly affect our results. The PV diagram essentially shows that the western filament is kinematically linked to the CND, with matching velocities at their interface, consistent with the CND being fed by the filament. The diagram also reveals a fast-rotating central component surrounded by a more disturbed, disk-like structure.

\subsection{Similarities to MHD simulations}\label{simsMHD}

To interpret the filamentary inflow and CND observed in NGC~4696, we compare our observations to tailored 3DMHD simulations designed specifically to reproduce the conditions of this system (see Section~\ref{sims}). These simulations build directly on the framework developed by \citet{Guo2023} and \citet{Guo2024}. Below, we first summarize the key physical processes identified in these simulations before comparing them directly to NGC~4696.

\citet{Guo2023} performed a set of nested-mesh hydrodynamic simulations with radiative cooling and heating that resolve the multiphase accretion flow from the event horizon of a SMBH in an elliptical galaxy out to galactic (kpc) scales. Building on this work, \citet{Guo2024} presented 3DMHD simulations in a similar framework, but including the impact of magnetic fields on the accretion flow. These simulations were designed to reproduce the turbulent cooling medium of a typical massive elliptical galaxy, specifically tailored to another central cluster galaxy, M87.

They found that relatively weak magnetic fields have a profound impact on the flow. In particular, the accretion rate increases by a factor of $\sim10$ when including magnetic fields.
What happens is an example of \textit{magnetohydrodynamic precipitation} \citep[see][for a tutorial]{Voit_2026arXiv260215121V}: magnetic-field tension can promote formation and precipitation of cold gas clouds by slowing the descent of cooling gas that is gaining density contrast \citep{Loewenstein_1990,Balbus_1991,Ji_2018,Wibking_2025MNRAS.544.2577W}. As the cooling gas descends, it vertically stretches and locally strengthens the magnetic field that is trailing it, shaping the cooling gas into a magnetized filament. The trailing magnetic tether that forms then applies a torque that drains the filament's angular momentum, so that it falls almost directly toward the center \citep{WangRuszkowskiYang_2020MNRAS.493.4065W}.

The simulated filaments therefore play a key role in fueling the SMBH. Large-scale Maxwell stresses provide one of the main mechanisms for angular momentum transport \citep{Guo2024}. As a result, the cold filaments rapidly lose angular momentum and rain back down toward the SMBH. However, rather than falling directly onto the black hole, the filamentary accretion first feeds a thick, cold, and highly magnetized CND with a typical size of a few tens of parsecs. This disk then dominates the final stage of accretion onto the SMBH \citep[see also][]{Gaburov2012}. Some filaments may also follow more complex conical or helical trajectories while raining back down toward the AGN \citep{Gaspari2013}.

The resulting inflow consequently exhibits three distinct regimes: (1) on scales of $\sim0.03$--$3$ kpc, cooling gas disorderly accretes along magnetized multiphase filaments; (2) on scales of $\sim0.3$--$30$ pc (or $10^3$--$10^5 \,r_g$), the flow transitions into a highly magnetized, rotating cold disk (essentially a CND); and (3) inside of $\sim0.3$ pc (or $10^3 \,r_g$), the flow becomes a magnetized, turbulent hot accretion flow. \citet{Guo2024} also note that, in addition to the accretion flow, strong outflows perpendicular to the CND are also driven by magnetic fields. Overall, the feedback from the SMBH outflows is sufficient to offset most of the cooling out to $\sim50$ kpc.

In Fig.~\ref{fig:simsobs}, we compare the CND observed in NGC~4696 with tailored 3DMHD simulations designed to reproduce the conditions of this system (see Section~\ref{sims}). The simulations show that gas cools and condenses into magnetized filaments (top left panel of Fig. \ref{fig:simsobs}) that channel material toward a disc around the central SMBH (top right panel of Fig. \ref{fig:simsobs}). The morphology bears a striking resemblance to the observations, with a similarly sized disk embedded within a network of filamentary structures. On larger scales ($\sim10 \mathrm{kpc}$), the simulations also exhibit a rich reservoir of cold gas, consistent with what is observed in NGC~4696 \citep{Crawford2005}. Most importantly, the simulations demonstrate that the filaments are not only kinematically linked to the disk, but continuously supply the CND, funnelling gas inward toward the SMBH. To enable a direct comparison to our data, we extract a PV diagram in the 3DMHD simulations along a slit extending from the center toward the top-right filament, mirroring the configuration used in the observations (see Fig.~\ref{fig:pv}). This comparison reveals a continuous velocity structure linking the filament to the CND, supporting a scenario in which filamentary gas flows inward and feeds the CND rather than being driven outward. This agreement provides strong support for a multiphase, filament-fed accretion process in which cold gas loses angular momentum mainly through Maxwell stresses, rains toward the black hole, and settles into a rotating CND before being accreted.

In the simulations, the disc size varies over time (right panel of Fig. \ref{fig:simsobs}), consistent with the behaviours reported in~\citet{Guo2023} and \citet{Guo2024}. We also show in Fig. \ref{fig:simsobs} the evolution of the orientation of the CND, as traced by its angular momentum along three different projection axes. The simulations demonstrate that filamentary accretion can induce significant variations in the orientation of the CND over time. If the angular momentum of the accretion flow is communicated to the central CND and then accretion disk, such CND wobbling could naturally lead to changes in the jet axis, providing an efficient mechanism for distributing AGN feedback heating more isotropically throughout the cluster core. We note that the present simulations still employ a morphological heating model, in which the heating rate balances cooling within each radial shell to maintain global equilibrium while permitting local thermal instability. A more realistic feedback prescription, as proposed in~\citet{Guo2025}, will enable the development of a self-consistent, multiscale model of the accretion flow that incorporates feedback-driven heating from first principles.


\subsection{Implications for AGN feedback}\label{imp}

In \citet{HL2025}, STIS/HST observations of the central SMBH in the galaxy cluster PKS 0745-191 ($P_{\rm cav}\sim5\times10^{45}$ erg s$^{-1}$; $\sim100\times$ more powerful than in NGC~4696) provided the first spatially resolved view of ionized gas dynamics within the sphere of influence under extreme feedback with $P_{\rm cav}>10^{45}$ erg s$^{-1}$. This is in contrast to NGC~4696, where $P_{\rm cav}\sim14^{+7}_{-4}\times10^{42}$ erg s$^{-1}$ \citep{Russell2013}. The kinematics in PKS 0745-191 appear disordered, with no evidence for a rotating CND, suggesting that strong feedback may disrupt CND formation and drive more disordered flows with stronger jet-gas coupling.

\citet{oosterloo2023} argued that the CND in NGC~1275 is not fully relaxed, with irregular outer kinematics consistent with ongoing filamentary accretion. Continuous inflow with varying angular momentum may reorient the CND, and if the inner jet remains perpendicular to the CND, this can drive changes in jet direction. Remarkably, a similar behaviour emerges naturally in our 3DMHD simulations (panel d) of Fig. \ref{fig:simsobs}), where ongoing accretion from filaments from various direction (see \ref{fig:sim_snapshots}) causes the CND to continuously change orientation with time. This filament-driven wobbling of the CND offers a plausible physical mechanism for reorienting AGN jets and may help explain how jet feedback can heat cluster cores in a nearly isotropic manner. A similar mechanism may operate in NGC~4696. In addition, sloshing motions in the cluster core \citep{XRISMNGC4696} may further perturb filament directions, leading to variations in the CND and jet orientation, which could in turn provide an efficient mechanism for isotropic heating in the cluster core over the lifetime of the cluster.

The strong similarity between our observations and the 3DMHD simulations in Figure \ref{fig:simsobs} points to a multiscale connection in the gas flow. As in \citet{Guo2024}, cold accretion \citep[e.g.,][]{Pizzolato2010,Li2012,Gaspari2013,Gaspari2017} arises from thermally unstable magnetothermal drip modes in the hot halo \citep[e.g.,][]{Loewenstein_1990,Balbus_1991,Ji_2018,Wibking_2025MNRAS.544.2577W,Voit_2026arXiv260215121V}, while magnetic fields with plasma $\beta\sim1-100$ organize the gas into thin filaments on kpc scales. This precipitation mechanism is morphologically consistent with the filaments observed in cluster cores, including NGC~4696, which require magnetic support \citep{Tamhane2026,Fabian2016,Sanders2008,Sanders2016,Fabian1982,Mittal2011,Canning2011,Crawford2005}; in NGC 1275, fields of $\sim24$ $\mu$G are sufficient \citep{Fabian2008,Tamhane2026}. These filaments can be extremely narrow ($\sim70$ pc) yet extend over $\sim6$ kpc. In NGC 1275, HCN(3-2) and HCO$^{+}$(3-2) absorption within 1.2 pc also reveal outflows of $\sim300$-$600$ km s$^{-1}$ which are consistent with simulations \citep{Guo2024}, though such features have yet been detected in NGC~4696.

Together, these results support a picture in which multiphase filaments transport gas from kpc to sub-kpc scales. Our observations suggest that the missing link is the formation of a CND: filamentary inflow feeds a cold CND that mediates accretion onto the SMBH, implying that Bondi accretion from hot gas is unlikely to dominate (see also Marquis et al. submitted). One key difference with simulations is that they predict predominantly cold CNDs \citep[where cold is defined as $2\times10^5$ K gas, see][]{Guo2024}, whereas here the CND is detected in Pa$\alpha$ ($\sim10^{4}$~K) and multiple H$_2$ transitions (a few $10^{2}$--$10^{3}$~K; Marquis et al., submitted), revealing a highly multiphase disk. This suggests that the disk is thermally stratified, with heated outer layers and colder gas embedded within clumps \citep{Gaspari2017}, a configuration that will be explored in detail in Pereira et al.\ (in prep.).


\section{Concluding remarks} \label{conc}

The growth of a galaxy’s central SMBH through accretion and the energetic feedback from the resulting AGN are tightly coupled processes that regulate galaxy evolution across vastly different scales. Understanding how this self-regulation operates remains a central challenge in astrophysics.

Our \emph{JWST}/NIRSpec observations provide a direct view of this connection. By resolving sub-kpc gas dynamics in NGC~4696, we show that the ionized swirl seen with \emph{HST} \citep{Fabian2016} is a compact CND physically connected to larger-scale filaments. This establishes the long-sought link between kpc-scale multiphase filaments and gas dynamics within $\lesssim100$ pc, where inflowing gas settles into a CND that feeds the SMBH. The rotating CND kinematics established here---a disk of radius $\sim120$~pc spanning $\sim600$~km~s$^{-1}$---provide the kinematic framework that will anchor the full multi-line analysis of Marquis et al.\ (submitted) and the ionization diagnostics of Pereira et al.\ (in prep.), enabling a complete multiphase picture of accretion in NGC~4696. The strong agreement with 3DMHD simulations indicates that this process arises naturally in magnetized atmospheres via thermally unstable magnetothermal drip modes. The later produce filaments with magnetic tethers capable of facilitating infall and accretion onto a rotating CND. By resolving this transition, our observations provide direct evidence that magnetized multiphase gas flows connect large-scale cooling to SMBH fueling, effectively closing the AGN feedback loop.

\begin{acknowledgments}

JHL acknowledges funding support from the Canada Research Chairs Program, as well as the Natural Sciences and Engineering Research Council of Canada (NSERC) through the Discovery Grant, Accelerator Supplement programs and the Arthur B. McDonald Fellowship. More importantly, most of this research was made possible through funding from the Canadian Space Agency (GRANT 24JWGO3A06). LA acknowledges support from the Canadian Space Agency (CSA) grant 22EXPJWST. MM acknowledges the support of the Natural Sciences and Engineering Research Council of Canada (NSERC) and of the Fonds de recherche du Qu\'ebec (FRQ) (doi:\href{https://doi.org/10.69777/365532}{10.69777/365532}). M Reefe acknowledges support from the National Science Foundation Graduate Research Fellowship under Grant No. 2141064. YL acknowledges support from NASA grant 80NSSC22K0668, Chandra X-ray Observatory grant TM3-24005X, NSF grants AST-2514692, AST-2510198, and CAREER award AST-2516092. J. B-B acknowledges support from project UNAM DGAPA-PAPIIT AG 101025, Mexico, as well as support from the PASPA 2025 grant.  HRR acknowledges support from an Anne McLaren Fellowship from the University of Nottingham and funding from a Leverhulme Trust Research Leadership Award.
PS acknowledges support by the French National Research Agency (ANR-25-CE31-5364). JSG thanks Macalester College for supporting the computational facilities used for this research. MLGM acknowledges financial support from NSERC via the Discovery grant program and the Canada Research Chair program.  MP acknowledges funding from the D\'{e}partement de Physique of the Universit\'{e} de Montr\'{e}al and the Centre for Research in Astrophysics of Quebec (CRAQ). B.V. acknowledges financial support from the D\'{e}partement de Physique of the Universit\'{e} de Montr\'{e}al. VO acknowledges support from DICYT ESO-Chile Comite Mixto PS 1757, and Fondecyt Regular 1251702. BRM acknowledges support from the Canadian Space Agency and the National Science and Engineering Research Council of Canada.

Some/all of the data presented in this paper were obtained from the Mikulski Archive for Space Telescopes (MAST). STScI is operated by the Association of Universities for Research in Astronomy, Inc., under NASA contract NAS5-26555. Support for MAST for non-HST data is provided by the NASA Office of Space Science via grant NNX13AC07G and by other grants and contracts.
The specific observations analyzed can be accessed via doi:\href{http://doi.org/10.17909/yqjb-gg89}{10.17909/yqjb-gg89}.

The authors are pleased to acknowledge that the simulation reported on in this paper was substantially performed using the Princeton Research Computing resources at Princeton University, which is consortium of groups led by the Princeton Institute for Computational Science and Engineering (PICSciE) and Office of Information Technology's Research Computing. 

The Universit\'e de Montr\'eal recognizes that it is located on unceded (no treaty) Indigenous territory, and wishes to salute those who, since time immemorial, have been its traditional custodians. The University expresses its respect for the contribution of Indigenous peoples to the culture of societies here and around the world. The Universit\'e de Montr\'eal is located where, long before French settlement, various Indigenous peoples interacted with one another. We wish to pay tribute to these Indigenous peoples, to their descendants, and to the spirit of fraternity that presided over the signing in 1701 of the Great Peace of Montr\'eal, a peace treaty founding lasting peaceful relations between France, its Indigenous allies and the Haudenosauni Confederacy (pronounced: O-di-no-sho-ni). The spirit of fraternity that inspired this treaty is a model for our academic community.

\end{acknowledgments}

\begin{contribution}

Authors 1-3 (J.H.-L. through M.G.) were instrumental in acquiring the observations and led the data reduction and analysis, simulations, writing, and interpretation. Authors 4-6 (M.M. through G.M.V.) contributed to the analysis and interpretation. All other authors (L.A. through S.W.) are listed alphabetically and contributed relevant scientific expertise to the project.

\end{contribution}

%
\facilities{HST(WFC3), JWST(NIRSpec), CXO.}

\software{astropy \citep{astropy:2013,astropy:2018,astropy:2022},  
          q3dfit \citep{2023ascl.soft10004R},
          GraphingLib \citep{gustave_coulombe_2026_19936793}, 
          NumPy \citep{harris_array_2020}, 
          pvextractor \citep{ginsburg_radio_2015}, 
          Matplotlib \citep{hunter_matplotlib_2007}.
          }


\appendix

\section{Temporal evolution of the CND in the 3DMHD simulations}

\begin{figure*}
\centering
\includegraphics[width = 0.96\textwidth]{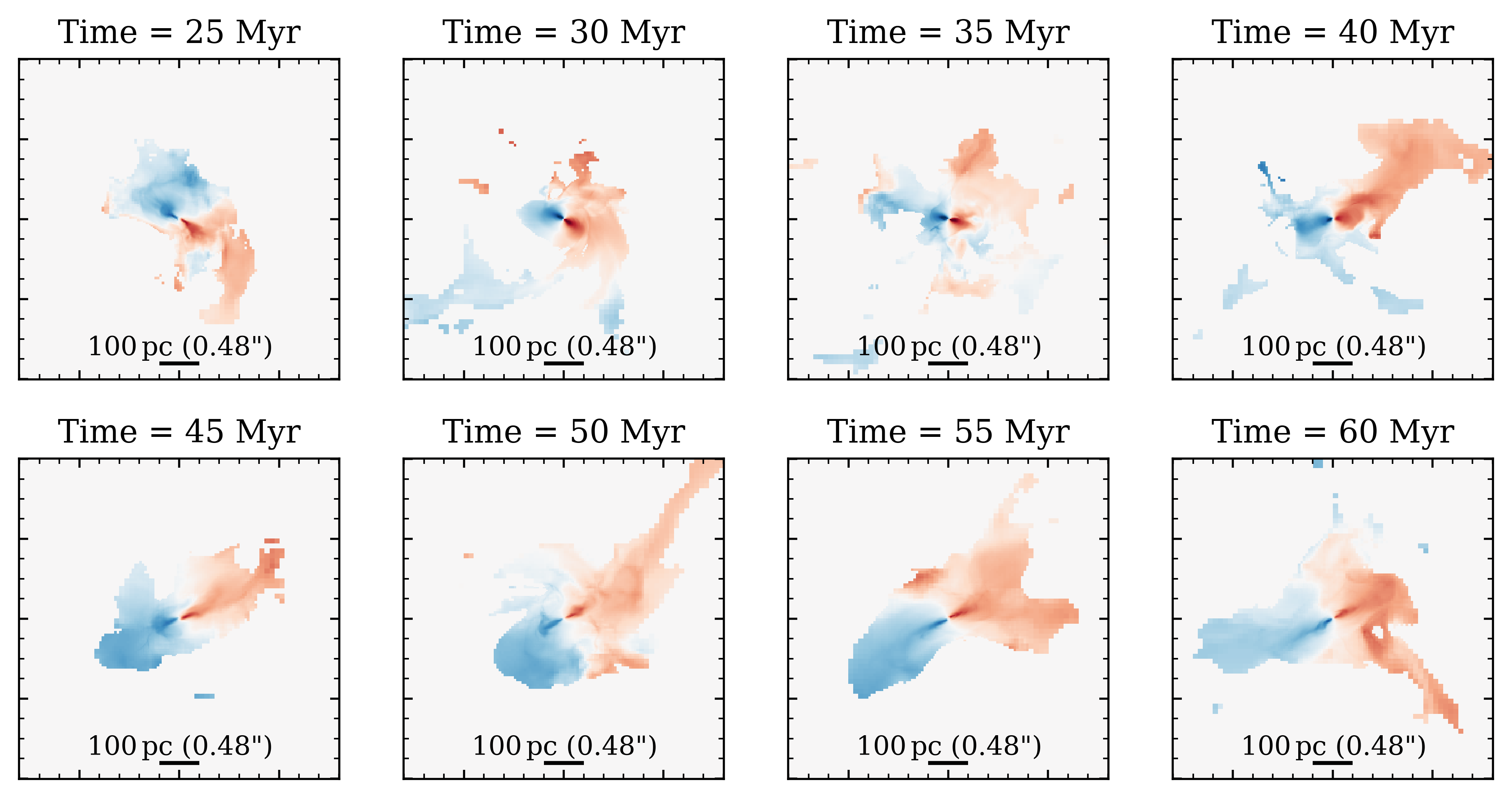}
\caption{Eight snapshots of the 3DMHD simulated velocity field within a JWST/NIRSpec IFU field of view, analogous to panel b) of Fig.~\ref{fig:simsobs}, and approximately evenly spaced in time. All panels are centered on the black hole position with the same surface density threshold in panel b) of Fig.~\ref{fig:simsobs}. The size and orientation of the CND vary with time, but the CND remains present and is typically connected to a larger-scale feeding filament.}
\label{fig:sim_snapshots}
\end{figure*}

Figure~\ref{fig:sim_snapshots} presents eight snapshots of the 3DMHD simulation, approximately evenly spaced in time and displayed over a field of view comparable to the JWST/NIRSpec observations. We omit the first $20$ Myr as it takes time for the CND to set up (see also panel c) of Fig. \ref{fig:simsobs}). All panels are centered on the black hole position. The morphology of the central gas distribution evolves substantially throughout the simulation, with the CND varying in both size and orientation (see also Fig. \ref{fig:simsobs}).

Despite this variability, a CND remains present in all snapshots. Furthermore, the CND is typically connected to one or more larger-scale filaments throughout the temporal evolution extending beyond the central few hundred parsecs. These filaments provide a direct pathway through which condensed multiphase gas can be transported toward the central regions.

The purpose of this figure is to demonstrate that the filament/CND connection shown in Figure~\ref{fig:simsobs} is not unique to a particular simulation output. Rather, the presence of a CND and its connection to larger-scale feeding filaments are common features throughout the evolution of the simulated system.

\bibliography{sample701}{}
\bibliographystyle{aasjournalv7}

\end{CJK}
\end{document}